\documentclass[12pt]{article}
\usepackage{amsmath}
\usepackage{epsfig}
\usepackage{amssymb}
\input{epsf}

\newcommand{\be}{\begin{equation}}
\newcommand{\ee}{\end{equation}}
\newcommand{\bea}{\begin{eqnarray}}
\newcommand{\eea}{\end{eqnarray}}

\newcommand{\vep}{\varepsilon}
\newcommand{\I}{{\cal I}}
\newcommand{\E}{{\cal E}}

\begin{document}
\def\sqr#1#2{{\vcenter{\hrule height.3pt
      \hbox{\vrule width.3pt height#2pt  \kern#1pt
         \vrule width.3pt}  \hrule height.3pt}}}
\def\square{\mathchoice{\sqr67\,}{\sqr67\,}\sqr{3}{3.5}\sqr{3}{3.5}}
\def\today{\ifcase\month\or
  January\or February\or March\or April\or May\or June\or July\or
  August\or September\or October\or November\or December\fi
  \space\number\day, \number\year}

\def\Bbb{\bf}
\topmargin=-0.3in

%%%%%%%%%%%%%%%%%%%%%%%%%%%%%%%%%%%%%%%%%%%%%%%%%%%

\newcommand{\ww}{\mbox{\tiny $\wedge$}}
\newcommand{\pp}{\partial}

\begin{flushright}
YITP-SB-06-23\\
%Draft\\
\today
\end{flushright}

\begin{center}
{\bf \Large NRQCD Factorization and Velocity-dependence of NNLO\\
\ \\
 Poles in Heavy Quarkonium Production}

\vbox{\vskip 0.45 in}

{\bf \large Gouranga C. Nayak$^1$, Jian-Wei Qiu$^2$, George Sterman$^1$}

\bigskip

{\it
$^{1}$ C. N. Yang Institute for Theoretical Physics, Stony Brook University
\\ Stony Brook, NY 11794-3840, USA \\
$^{2}$ Department of Physics and Astronomy, Iowa State University,
Ames, IA 50011, USA  }
\end{center}

\bigskip

\begin{abstract}
We study the transition of a heavy quark pair from octet to singlet 
color configurations at next-to-next-to-leading order (NNLO)
in heavy quarkonium  production.  We show that
the infrared singularities in this process are consistent with
NRQCD factorization to all orders in the heavy quark relative
velocity $v$. This factorization requires the gauge-completed
matrix elements that we introduced previously to prove NNLO factorization
to order $v ^2$.    
\end{abstract}

\section{Introduction}

Heavy quarkonium production serves as a testing ground for 
perturbative and effective field theory treatments of
QCD, particularly nonrelativistic QCD (NRQCD) \cite{nrqcd}.
NRQCD, which relies on an expansion in the pair 
relative velocity, as well as in $\alpha_s$, has
provided compelling explanations for 
quarkonium production at 
collider \cite{collider} and fixed target experiments \cite{fixed}.
At the same time, puzzles remain, especially
from polarization measurements at the
Tevatron \cite{polarization}, and associated production at the
B factories \cite{associated}. 
Heavy quarkonium 
production in the evolving QCD medium at the RHIC and LHC \cite{qgp}
may also play a role in the detection 
and analysis of new states of strongly-interacting matter.
For recent updates on heavy quarkonium production at zero and finite temperature
QCD see Refs.\ \cite{cdf, review}.

The application of NRQCD to heavy quarkonium production processes
is based on a very specific factorization property \cite{nrqcd},
for which a 
complete proof has not yet been developed \cite{review}.
With this in mind, we
tested the factorization hypothesis at 
next-to-next-to-leading order (NNLO) in Ref.\ \cite{fact2}.  We found
that, as a necessary condition for factorization, the conventional ${\cal O}(v^2)$
octet NRQCD  production matrix elements must be 
redefined by incorporating Wilson lines that make them manifestly gauge invariant.
We referred to these as gauge-completed matrix elements.

This result was derived by employing an
eikonal approximation for the coupling of soft
radiation to the heavy quarks.  
To order $v^2$, this approximation 
is adequate to treat the lowest-order
electric dipole transitions \cite{braatenreview}
that transform an octet pair to a singlet,
without modifying the spin.
In this paper, we extend our eikonal NNLO analysis,
and show that the factorization of
such transitions 
can be extended from order $v^2$ to finite $v$, that is, to
all orders in the relative velocity,
in terms of gauge-completed NRQCD matrix elements.  
The essential result is that at all orders in $v$,
the infrared divergence at NNLO is independent
of the direction of the light-like Wilson line
that renders the matrix element gauge invariant.
We find this result intriguing, and while
this modest extension of our previous
result does not address the behavior
of spin-dependent operators, it should
encourage further work on the factorization theorem.

In the following section we review the role of gauge-completed
matrix elements
in NRQCD factorization, and the requirements that 
factorization places on these matrix elements and the
infrared poles in dimensional regularization that they must match.
We also introduce the specific NNLO eikonal factor that
we will calculate to test factorization at this order, and
give the result of our calculations.
The details of our NNLO calculation are presented in
Sec.\ 3, where we verify that the necessary conditions
for NRQCD factorization are met to all
powers in the relative velocity, after which we
give a brief conclusion.

\section{Infrared Poles and NRQCD Factorization}

In NRQCD, 
the production cross section for heavy quarkonium $H$
at transverse momentum $p_T$
factorizes into a sum of perturbative functions times 
universal matrix elements,
\begin{equation}
d\sigma_{A+B\to H+X}(p_T) = \sum_n\; d\hat\sigma_{A+B\to 
c\bar{c}[n]+X}(p_T)\,
\langle {\mathcal  O}^H_n\rangle\, ,
\label{nrfact}
\end{equation}
where each matrix element $\langle {\mathcal O}^H_n\rangle$ 
represents the
probability for a heavy quark pair in state $[n]$ 
to produce quarkonium  state $H$.  
The states $n$ may,
in particular, be color octet or singlet.

Correspondingly,
at large $p_T$,
the fragmentation function for
parton $i$ to evolve into a  heavy quarkonium  is factorized according to 
  \cite{braatenreview}
\begin{equation}
D_{H/i}(z,m_c,\mu) = \sum_n\; d_{i\to c\bar{c}[n]}(z,m_c,\mu) \, 
\langle  {\mathcal O}^H_n\rangle\, ,
\label{combofact}
\end{equation}
in terms of the same matrix elements,
along with
perturbative functions $d_{i\to c\bar{c}[n]}(z,m_c,\mu)$ that
describe the evolution of an 
off-shell parton into a quark pair in state $[n]$, including logarithms of 
$\mu/m_c$.

Although we cannot compute the full fragmentation function
in perturbation theory, 
for NRQCD factorization to hold
we must be able to compute
the function $d_{i\to c\bar{c}[n]}(z,m_c,\mu)$ in Eq.~(\ref{combofact})
systematically.  To do this, we must be able to
use regularized perturbation theory to compute probabilities
for the production of a singlet quark pair from 
any local source, and match the infrared singularities
of these probabilities to the matrix elements
in the sum over $n$ in Eq.~(\ref{nrfact}) and/or Eq.~(\ref{combofact}).  

\subsection{Gauge-completed matrix elements}

As advocated in \cite{fact2}, it is natural  to 
define gauge-invariant
octet NRQCD operators ${\cal O}^H_n$, of the general form
\bea
\left \langle \, {\mathcal O}^H_n(0)\, \right \rangle
=
\left \langle \, 0\, \left |\,
\chi^\dagger(0){\mathcal \kappa}_{n,e}\psi(0)\, 
\Phi_l^{(A)}{}^\dagger(0){}_{eb}\, 
\left(a^\dagger_Ha_H\right)\,
\Phi_l^{(A)} (0)_{ba}\, \psi^\dagger(0) {\mathcal \kappa}'_{n,a}\chi(0)
\,  \right|\, 0\, \right \rangle\, ,
\label{replace}
\eea
in terms of  heavy quark ($\psi$) and antiquark ($\chi$) operators, and
local combinations of color and spin matrices and/or covariant derivatives,
denoted by $\kappa_{n,e}$ and $\kappa'_{n,a}$.
The operator $a^\dagger_H$ creates quarkonium $H$, and
the operators $\Phi^{(A)}_{\ell}(0)$ are
Wilson lines, that is, ordered exponentials, constructed
from the gauge
field in adjoint matrix representation, $A^{(A)}_\mu$, as 
\bea
\Phi^{(A)}_l(0) = {\cal P} \exp 
\left[ -ig\int_0^\infty d\lambda\, l\cdot A^{(A)}(l\, \lambda)\right]\, ,
\label{oexp}
\eea
where $\cal P$ denotes path ordering
and $\ell^\mu$ is the velocity  of the source.  
In (complex conjugate) amplitudes, (anti)time-ordering
is understood.  

For NRQCD factorization to hold,
a necessary, and superficially paradoxical, property of
the gauge-completed matrix elements  is that 
their long-distance behavior 
must be independent of the vector $l^\mu$ that we choose to
define them \cite{fact2}.   Such a dependence would
be inconsistent with NRQCD factorization, because
the infrared divergences of ${\mathcal O}^H_n$
must match those of cross sections, in which
there is no information on $l^\mu$.  
In Ref.\ \cite{fact2}, we have verified
the $l$-independence
of the infrared pole
 to order $v^2$ in the
relative velocity of the pair, at order NNLO.
We will extend this result below, to all powers
in $v$, again at NNLO.

\subsection{Matrix elements and infrared universality}

In this paper, we will study the infrared 
behavior of the octet, S-wave matrix elements
\bea
{\cal M}^{(8\to I)}(P_1,P_2,l) =
\sum_X\, \langle 0|\, \chi^\dagger(0)\, T^{(q)}_e\psi(0)\, 
\Phi_l^{(A)}{}^\dagger(0){}_{eb}\,  |[c(P_1)\bar{c}(P_2)]^{(I)}\; X\rangle
\nonumber\\ 
 \times
\langle X\, [c(P_1)\bar{c}(P_2)]^{(I)}\, |\,
\Phi_l^{(A)} (0)_{ba}\, \psi^\dagger(0) \, T^{(q)}_a\chi(0)\,
|0\rangle\, ,
\label{calM}
\eea
where $I=1,8$ labels the color of the heavy quark pair,
with momenta $P_1$ and  $P_2$.  The $T^{(q)}_i$ are
color generators in the quark fundamental representation.
We will compute the infrared poles of Eq.~(\ref{calM}) 
to NNLO.
At lowest order of
course, only $I=8$ contributes in the final state.  
At higher orders, however,
the octet state mixes with the singlet state, through
radiation to states $X$.  The phase space of $X$
is cut off at a UV scale $\mu$, which we take to
be of the order of the heavy quark mass.

In Eq.\ (\ref{calM}),
we specify the pair's momenta in conventional fashion, 
in terms of the total pair ($P$) and  relative momentum ($q$),
and center-of-mass velocity ($v$) by
\bea
P_1 = \frac{P}{2} + q\, , \quad \quad 
P_2 = \frac{P}{2} - q\, , \quad \quad 
v^2=\frac{\vec{q}{\, }^2}{{E^*}^2}\, ,
\label{P1P2Pq}
\eea
where $2E^*$ is the total center of mass energy of the pair in the
pair's rest frame.  We note that our convention for the
relative
 velocity here differs from the one in our previous study, Ref.\ \cite{fact2},
where we defined the relative velocity as $2\vec q/m$, with $m$
the heavy quark mass.  

NRQCD requires that when we expand $\cal M$ of Eq.\ (\ref{calM}) 
in powers of $q$ in perturbation theory, 
we should  find infrared-finite coefficient functions times
infrared-sensitive but universal NRQCD matrix elements,
\bea
{\cal M}^{(8\to I)}(P_1,P_2,l) = \sum_n \hat{\cal C}^{(8)}_n(P_1,P_2,l)\,
\langle  {\mathcal O}^I_n\rangle \, ,
\label{calMfact}
\eea
where $\hat{\cal C}^{(8)}_n$ is a perturbative coefficient function 
for NRQCD operator ${\mathcal O}^I_n$
with the sum over all operators $n$ for a given final state
in which the heavy quark pair has color $I$.
To test these ideas, we must study the infrared behavior
of such a matrix element when the color of the final state is fixed
as a singlet, $I=1$.

 The formation of a heavy quarkonium state, of course,
 cannot be realized to any fixed order of perturbation theory.
Nevertheless, the factorization expressed
in Eqs.\ (\ref{nrfact}), (\ref{combofact}) and (\ref{calMfact}) is useful
to the extent that we can systematically compute corrections
to short-distance functions in 
each case.   This, in turn, requires that the infrared poles
encountered in the evolution of the heavy quark pair 
from octet to singlet in $\langle {\cal O}_n^H\rangle$
in Eq.~(\ref{combofact}) match those of cross sections for the
production of  a singlet quark pair.
As already noted, a necessary condition
for this matching is that the poles in the matrix element should
not depend on the direction of the vector $l^\mu$ that is 
introduced in gauge completion of the matrix elements, Eq.\ (\ref{replace}),
since the choice of $l^\mu$ is matter of convention.

Now we  are ready to describe the NNLO calculation
that tests these ideas.  
A full NNLO calculation of ${\cal M}^{(8\to 1)}$
in Eq.\ (\ref{calM}) would be  impractical, but 
its infrared singularities are easier to compute.
These divergences can be generated 
by a factorization that is much simpler than Eq.\ (\ref{calMfact}),
and which can be carried out at fixed momenta $P_1$ and $P_2$,
\bea
{\cal M}^{(8\to 1)}(P_1,P_2,l) =
\sum_J {\mathcal C}_{8J}(P_1,P_2,l)\; {\cal E}^{(J\to 1)}(P_1,P_2,\vep) 
\, .
\label{schemefact}
\eea
All spin information in ${\cal M}^{(8\to 1)}$ is contained in another
short-distance function ${\mathcal C}_{8J}$,  which describes
a transition of octet to color  configuration $J$
at short distances.  Although 
${\mathcal C}_{8J}$ may depend on $l^\mu$, it
must be finite for $\vep \to 0$.
All $1/\vep$ poles are absorbed into 
an infrared factor, ${\cal E}^{(8\to 1)}$,
whose pole structure,
however, must be independent of $l^\mu$.
The expansion of the 
infrared factor,
${\cal E}^{(J\to 1)}(P_1,P_2,\vep)$ 
in the relative velocity, $v$ of the pair should lead us back
to a set of $l$-independent operator matrix elements for producing a color singlet
quark pair ($I=1$) in Eq.\ (\ref{calMfact}).
The essential result of this paper is that at NNLO the 
infrared poles of the function
${\cal E}^{(8\to 1)}(P_1,P_2,\vep)$ are indeed independent of
the vector $l$, for arbitrary $v$.   This generalizes the result
of  Ref.\ \cite{fact2} from 
$v^2$ ``electric dipole" transitions to arbitrary powers of 
$v$ at NNLO.

Specifically, we 
will find an explicit single-pole contribution, which can be written as a 
prefactor determined by $N_c$, the number of colors, times a
function that depends only on the relative velocity,
\bea
{\cal E}^{(8\to 1)}(P_1,P_2,\vep) 
&=&
 -\ \frac{N_c}{4}\, \left(N_c^2-1\right)\, \I^{8\to 1}(v,\vep)\, .
\label{Fcolor}
 \eea
The velocity-dependent factor $\I^{8\to 1}(v,\vep)$ is given by
 \bea
{\cal I}^{(8\to 1)}(v,\vep) 
&=&
 \, \frac{~\alpha_s^2}{4 \vep} \; 
\left \{1-~\frac{1}{2 f(|\vec{v}|)}\; 
 \ln \left[\frac{1+ f(|\vec{v}|)}{1- f(|\vec{v}|)}\right] \right \}\, ,
\label{fnv}
\eea
where
 ($v$ = $|\vec{v}|$) is the relative velocity of
the quark and antiquark in 
the pair
center of mass,
and where $f(v) = 2v/(1+v^2)$. As anticipated, $\I^{8\to 1}(v,\vep)$
is independent of $l$.
We will compute the color factor
given
in Eq.~(\ref{Fcolor}) below.
To derive the result of Eq.\ (\ref{fnv}), we will first recall the use of the 
eikonal approximation to isolate infrared behavior.

\subsection{Soft gluon interactions}

The eikonal apporoximation
 reproduces all infrared divergences in the evolution
 of the pair into the final state, and it must be defined
 by an infrared regularization.  We will
use a continuation to $4-2\vep$ dimensions, with $\vep<0$.

  The eikonal approximation
  for the interactions of the heavy quarks with  soft gluons is 
 generated  by ordered exponentials,
  this time in fundamental representations and in
  the directions of the  heavy quark and antiquark momenta.
   The perturbation theory rules for the
 ordered exponentials are equivalent to the eikonal approximation.
 Eikonal quark propagators and gluon-quark
  vertices are specified respectively by
   \bea
  \frac{i}{(\beta\cdot k+i\epsilon)} \quad  ,
  \quad
  \pm ig_sT_a^{(q)}\beta^\mu\, , 
 \label{eikrules}
 \eea
where the plus is for the antiquark and the minus for the quark
vertices,  and where
$\beta^\mu$ is a time-like 
four-velocity.    
Because the product of an eikonal propagator and
 vertex is always scale invariant, we will use
 below the momenta defined in Eq.~(\ref{P1P2Pq}) for the eikonal
 velocities of the quark pair.  

The long-distance evolution of
the pair from octet to singlet color configurations
is given by the 
infrared
factor ${\cal E}^{(8\to 1)}$
of Eq.\ (\ref{schemefact}),  which can be represented as
a matrix element.
This matrix element is given in the notation
  of Eq.\ (\ref{calM}) by
  \bea
{\cal E}^{(8\to 1)}(P_1,P_2,\vep)
&=& 
\sum_N\, 
\langle 0|\, \left [\Phi_{P_2}^{(\bar{q})}{}^\dagger (0)\right]_{IJ} 
\left[T^{(q)}_e\right]_{JK}\ 
\left[\Phi_{P_1}^{(q)}{}^\dagger(0) \right]_{KI}\, 
\Phi_l^{(A)}{}^\dagger (0)_{eb}\, \left| N\right\rangle
\nonumber\\
&\, & \hspace{5mm} \times
\left< N \right| 
\Phi_l^{(A)} (0)_{ba}\, 
\left[\Phi_{P_1}^{(q)}(0)\right]_{LM} \left[T^{(q)}_a\right]_{MN} 
\left[ \Phi_{P_2}^{(\bar{q})}(0)\right]_{NL}
\, |0\rangle \, .
\label{Edef}
\eea
Here we have exhibited all color indices: those in
adjoint representation by $a,\, b \dots$, and those in
the fundamental representation by $I,\, J\dots $, to indicate
the trace structure, which imposes a color singlet configuration
on the quark pair in the final state.  
In Eq.\ (\ref{Edef}) and below,
overall time-ordering of the field operators is understood in the amplitude,
and anti-time ordering in its complex conjugate.

The operators  $\Phi^{(q)}$ and $\Phi^{(\bar q)}$ are the
ordered exponentials that represent the quark and antiquark.
The quark (antiquark) ordered exponential
 has the same (opposite) ordering compared
to time ordering.  To be specific,  we represent
normal (reverse) matrix ordering  by ${\cal P}$ ($\bar {\cal P}$), and define 
\bea
\Phi_{P_1}^{({q})}(0)
&=&  {\cal P} \exp \left[ - ig\int_0^\infty d\lambda\,  
P_1\cdot A^{(q)}(P_1\lambda)\right]\, ,
\nonumber\\
\Phi_{P_2}^{(\bar{q})}(0)
&=& \bar{\cal P} \exp \left[ ig\int_0^\infty d\lambda\,  
P_2\cdot A^{(q)}(P_2\lambda)\right]\, .
\label{Phiqbar}
\eea
For classical fields, $\Phi_n^{(\bar q)}(0)$ is the  hermitian conjugate
of $\Phi_n^{(q)}(0)$.  
The matrix   $A^{(q)}_\nu \equiv \sum_a T^{(q)}_a\, A_{\nu,\, a}$ is the
gauge field operator in the quark fundamental representation.

The matrix element in Eq.~(\ref{Edef}) is equal to unity
when $q=0$.
 In Ref.\ \cite{fact2} we expanded Eq.\ (\ref{Edef}) to  
second order in  $q$, and found an expression in terms
of field strength operators,
     \bea
    {\cal E}_2^{(8\to 1)}(p+q,p-q,\vep) &\equiv&
    \sum_N\, 
     \int_0^\infty d\lambda' \,  \lambda' \, \left< 0 \right| \,  
       \Phi_l^{(A)}{}^\dagger (0)_{bd'}\,
     \Phi_p^{(A)}(\lambda'){}^\dagger_{d'a'}\,
        \left[\, p^\mu q^\nu F_{\nu\mu,a'}(\lambda' p)\, \right]\,
       \left|  N\right\rangle
       \nonumber\\
       &\ & \hspace{5mm} \times  \left\langle N\right|
    \int_0^\infty d\lambda \,  \lambda \,  \Phi_l^{(A)} (0)_{bd}\,
   \left[\, p^\mu q^\nu F_{\nu\mu,a}(\lambda p)\, \right]
     \Phi_p^{(A)}(\lambda)_{ad}\, \, |0\rangle\, .
            \label{I2def}
       \eea
        In this expression, the momenta $p^\mu$ and $q^\mu$ are taken to be dimensionless,
 scaled by the heavy quark mass $m$.
We notice that in the heavy quark rest frame, 
$p \equiv (1/2m)(P_1+P_2)_{\rm rest} = \delta_{\mu 0}$,
the relevant operator is precisely the chromo-electric field $F_{\mu 0}$,
and the matrix elements describe an electric dipole transition.

The basic result of Ref.\ \cite{fact2} was to identify an infrared
pole in 
${\cal E}_2^{(8\to 1)}$, associated with the exchange of gluons
between the heavy quark pair and the eikonal source 
$\Phi_l$.  The presence of such a pole showed
first that the infrared behavior of the fragmentation
function is not summarized by ``topologically factorized"
diagrams alone in an arbitrary gauge \cite{braatenreview}.  The specific form, however, depends on
the relative velocity of the pair only.  In terms of the center-of-mass velocity
of the heavy quark and antiquark, defined as in Eq.~(\ref{P1P2Pq}) above,
the result is \footnote{The coefficient of $v^2$ here is four times larger 
than in Ref.\ \cite{fact2}, where the velocity was defined by $2\vec q/m$,
as noted in connection with Eq.\ (\ref{P1P2Pq}) above.}
\begin{equation}
%% I -> 1
{\cal E}_2^{(8\to 1)}(v)~
~=~ \ \frac{N_c}{4}\, \left(N_c^2-1\right)\, \alpha_s^2~\frac{1}{3 \varepsilon}~v^2\, .
\label{svq}
\end{equation}
The prefactor here is the 
same color factor given in Eq.\ (\ref{Fcolor}) above.
The crucial point is that the pole due to 
topologically non-factored diagrams is independent of the direction of the
light-like vector $l^\mu$. In Ref.\ \cite{fact2}
we showed that this implies that the gauge-completion
of the NRQCD matrix element is adequate to 
match this 
infrared structure in production processes
with arbitrary numbers of recoiling jets.  This verified the NRQCD 
factorization of Eq.\ (\ref{combofact}) at NNLO in soft gluon corrections 
for octet operators.
We will show here that this result generalizes to all
orders in $v$ for the full infrared factor 
${\cal E}^{(8\to 1)}$.

In our calculation below, we will keep $q$, or equivalently
the relative velocity $v$, finite and nonzero, and evaluate Eq.\ (\ref{Edef})
directly to NNLO ($\alpha_s^2$), without an expansion in 
velocity.  Restricting ourselves to this order in $\alpha_s$,
an expansion in $v$ will be describable at any order
in terms of derivatives of the electric field strength.
In this sense our calculation identifies the infrared pole
in the sum of electric multipole transitions generated by
NNLO.   Because of the eikonal approximation,  no information
on spin-dependence is included in the calculation,
and the consequences for matching are similar to 
the $v^2$ case.  We will once again find an infrared
pole to any order in $v^2$, and we will once again find
that the pole is independent of the direction of the
vector $l^\mu$.  This surprising result shows that
matching to gauge-completed NRQCD matrix elements
is not limited to lowest order in $v^2$, but to this
order in $\alpha_s$ and for this class of electric dipole transitions,
is true to all orders in the relative velocity $v$.

\section{Finite-$v$ diagrams}

\subsection{Ladder and three-gluon diagrams}

Our goal is to calculate the non-canceling
infrared pole term in Eq.\ (\ref{Edef}),
the  eikonal infrared factor at finite 
relative velocity $v$.   As in Ref.\ \cite{fact2}, we need
only consider diagrams that are not topologically
factorized.   As indicated above, we take the
quark momentum as $P_1$, the antiquark as $P_2$,
and the momentum of the  gauge line as $l$, with
$l^2=0$.   The gauge line may be thought of as 
part of a fragmentation function, or as representing
an approximation to a recoiling jet in a hard-scattering
cross section.  
\begin{figure}
\begin{center}
\epsfig{figure=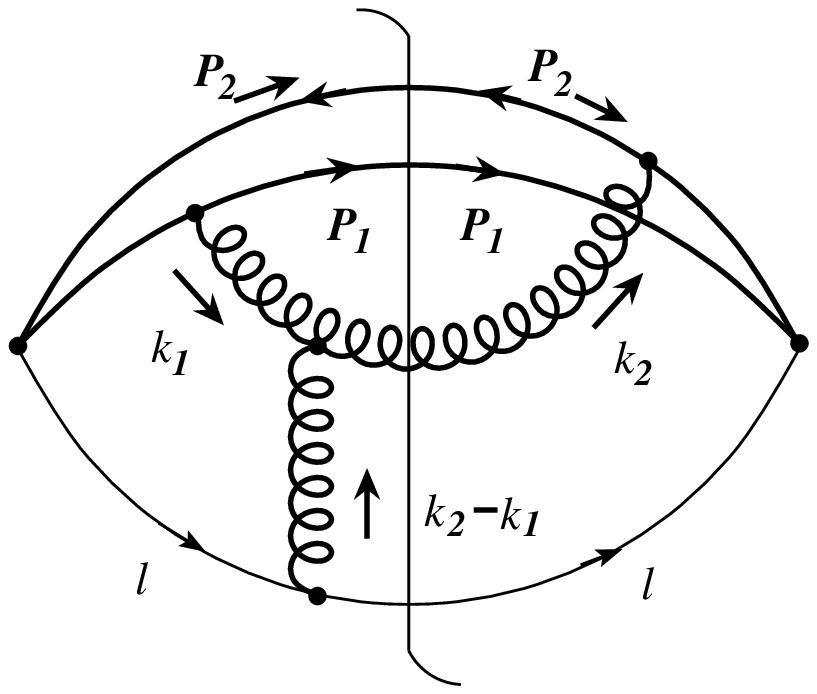,width=0.2\textwidth}
\hfil\\
(a)
\vskip 5mm
\epsfig{figure=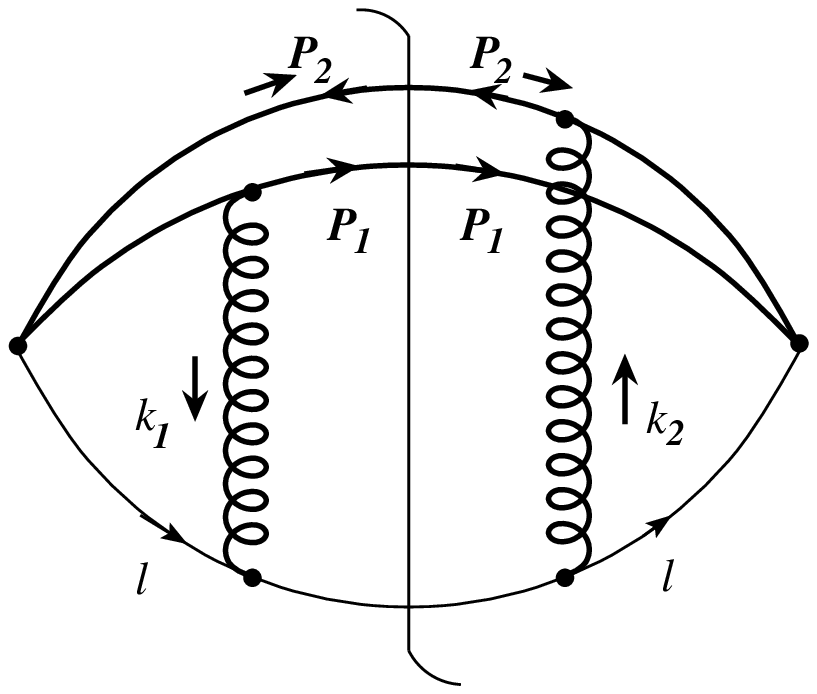,width=0.2\textwidth}
\hfil
\epsfig{figure=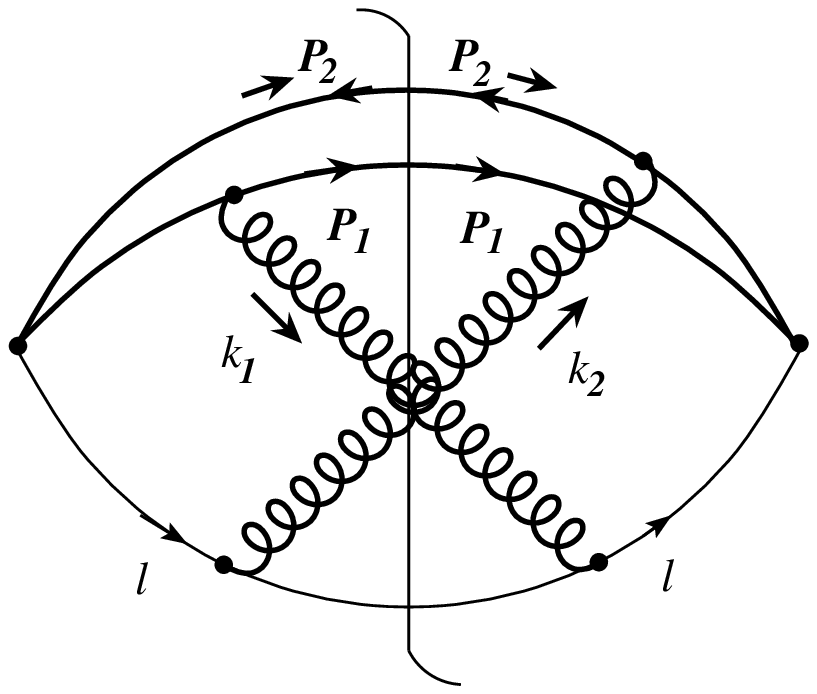,width=0.2\textwidth}
\hfil
\epsfig{figure=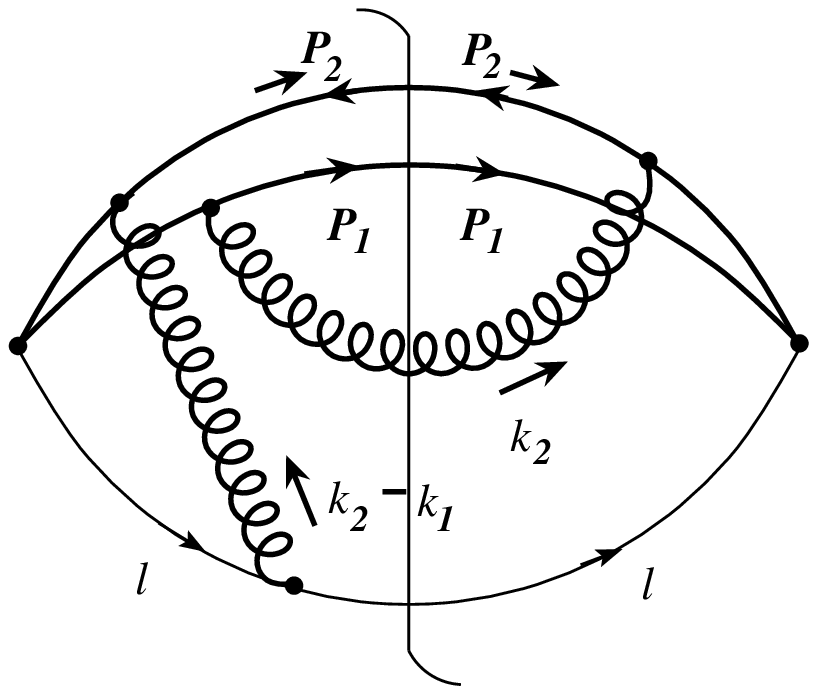,width=0.2\textwidth}
\hfil\\
(b)
\end{center}
\caption{Examples of NNLO diagrams for the quarkonium
production. 
 a) Three-gluon diagram b) 
QED-like diagrams. The projection of the quark pair onto a color singlet in
 the final state is understood. \label{diagrams}}
\end{figure}

The relevant diagrams are illustrated in Fig.\ \ref{diagrams},
with quark-antiquark eikonal lines on top, and the
gluon eikonal below.   
In each case, the vertical line indicates
the color-singlet heavy quark pair, and the sum over other
possible final states 
is understood.
The left- and right-most vertices in each diagram couple the light-like color
octet eikonal on the bottom of the diagram
with the color octet projection of the eikonals
on the top that represent the heavy quark pair,
in the fundamental representations.  

The quark lines radiate soft gluons to transform
themselves from color octet to color singlet, as in the
classic description of the color octet mechanism.
At NNLO, however, these soft gluons can scatter
from the eikonal gluon in the $l$ direction.  
To this order, the diagrams either contain the three-gluon coupling
as in Fig.\ \ref{diagrams}a, or are QED-like
as in Fig.\ \ref{diagrams}b.
Of these, only the former give rise to  noncanceling 
infrared poles, generalizing the result of 
Ref.\ \cite{fact2}. \footnote{It is worth noting that the color factor
of the left-most diagram in Fig.\ \ref{diagrams}b vanishes in
the matrix element of Eq.\ (\ref{Edef}).}
The color factor in Eq.\ (\ref{fnv}) is thus found entirely from
Fig.\ \ref{diagrams}a.   
This factor is the product of two simple traces in fundamental representation,
and the color tensors from two three-gluon vertices,
\bea
F_{\rm color} &=& {\rm Tr}\left[ T^{(q)}_e\, T^{(q)}_g\,  \right] \, 
{\rm Tr}\left[ T^{(q)}_a\, T^{(q)}_h\,  \right] \, 
f_{eai}\, f_{ihg}\, ,
\nonumber\\
&=& -\ \frac{N_c}{4}\, \left(N_c^2-1\right)\, .
\label{Fcolorcalc}
\eea
The two traces reflect the projection of the quark-antiquark
eikonal pair onto the
singlet in the final state that is built into the
matrix element of Eq.\ (\ref{Edef}). 
In the traces, one generator is associated with 
the operator in the
corresponding
matrix element and the other 
with
the corresponding quark-gluon vertex shown in Fig.\ \ref{diagrams}a. 
 We now turn to the  calculation of the 
velocity dependence of ${\cal E}^{(8\to 1)}$.

\subsection{Velocity-dependence of  ${\cal  E}^{(8\to 1)}$}

\begin{figure}[h]
\begin{center}
\epsfig{figure=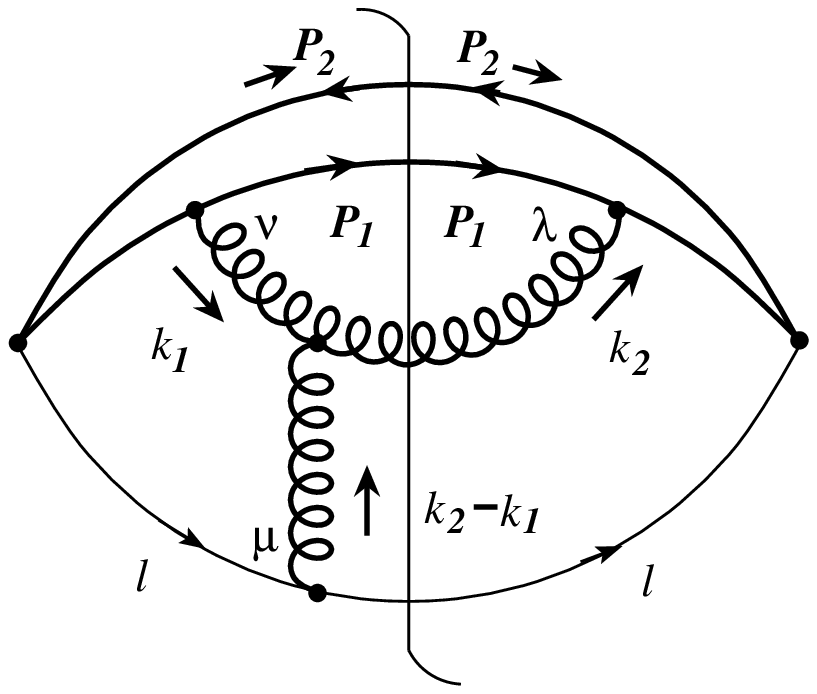,width=0.2\textwidth}
\hfil
\epsfig{figure=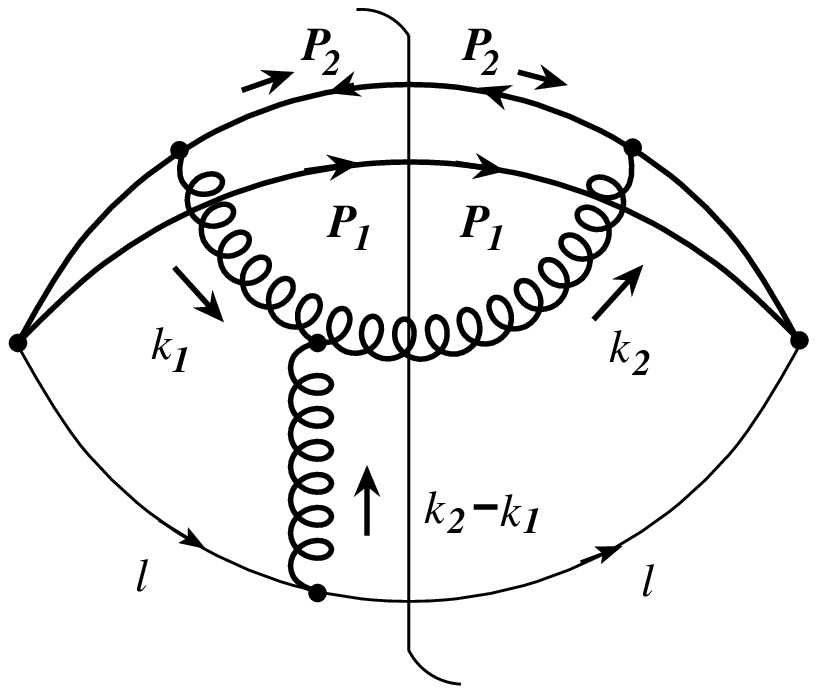,width=0.2\textwidth}
\hfil
\epsfig{figure=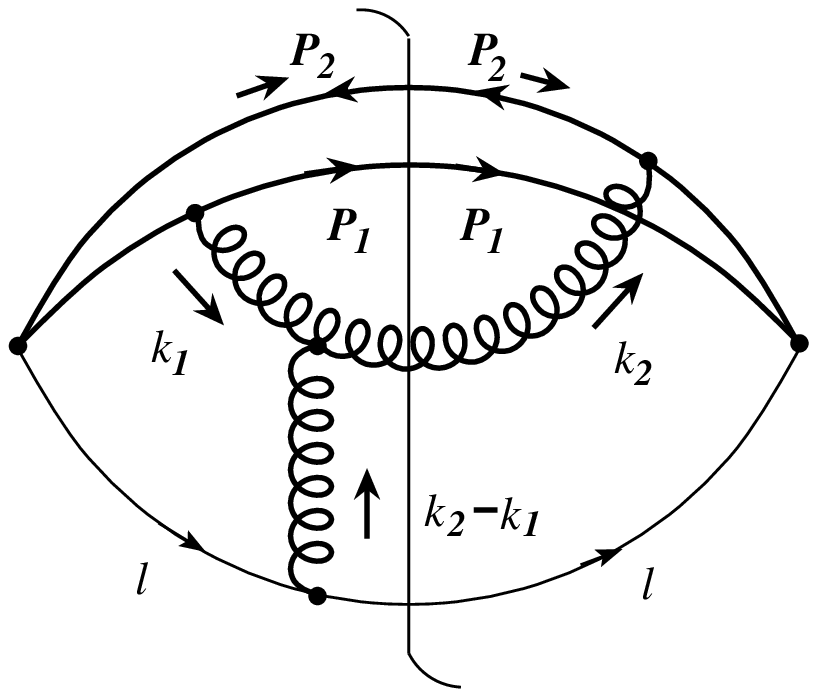,width=0.2\textwidth}
\hfil
\epsfig{figure=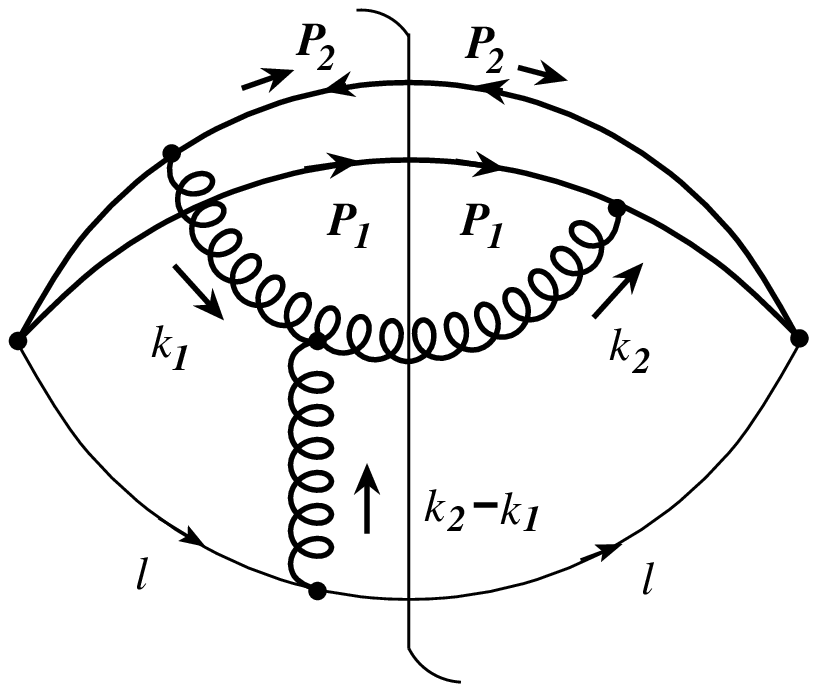,width=0.2\textwidth}
\hfil\\
(a)
\vskip 8mm
\epsfig{figure=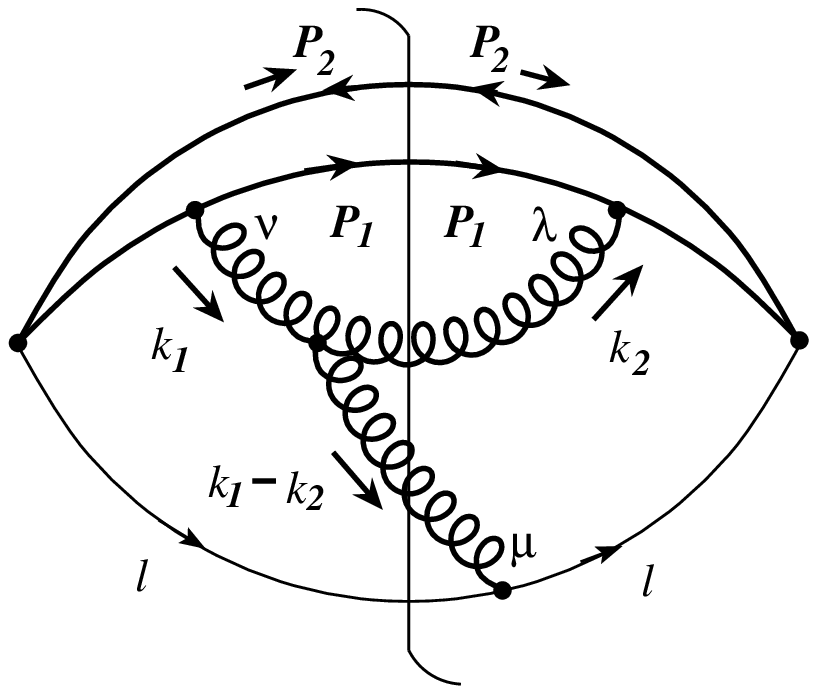,width=0.2\textwidth}
\hfil\\
(b)
\end{center}
 \caption{a) The four relevant cut diagrams with three-gluon vertices.
 The $P_1$, $P_2$ and $l$ lines are eikonal.
 The full contribution of each diagram to $\I^{(8\to 1)}$ is
 found by summing over other cuts that include the
 singlet pair. b) An example of a two-gluon final state
 that cancels the $(k_2-k_1)^2=0$ pole, as discussed in the text.
 As in Fig.\ 1, the projection of the quark pair onto a color singlet in
 the final state is understood.
 \label{fourdiagrams}}
  \end{figure}

We will give a detailed discussion of the cut
diagrams  illustrated in Fig.\ \ref{fourdiagrams}.
 In these four cut diagrams,  a soft gluon is emitted by
 one member of the (eikonal) heavy quark pair in the
 amplitude and is absorbed by one in the complex conjugate amplitude.  
 This soft gluon rescatters from the spectator eikonal gluon
 of momentum $l$ in either the amplitude or complex conjugate. 
 For each of these diagrams
 the leading order $v^2$ expansion of this contribution to 
$\E_2$, Eq.\ (\ref{I2def}),
 was calculated in detail in Ref.\ \cite{fact2} (where
 it was referred to as diagram IIIA).
 Here we extend this calculation to all orders in $v$
  by directly calculating the full velocity dependence of
$\E^{(8\to 1)}$.

 All of the diagrams in Fig.\ 2 have the same color factor
 (the same as in Eq.\ (\ref{Fcolorcalc}) above) when the quark pair
 is projected onto a color singlet in the final state. We will
 suppress the common color factor in our calculation
 below, and denote the sum of the relevant NNLO diagrams
 as 
$\I^{(8\to 1)}={\cal E}^{(8\to 1)}/F_{\rm color}$.
We label the diagrams of Fig.\ \ref{fourdiagrams},
whose sum gives $\I^{(8 \to 1)}$, by
\bea
&& \I^{(8\to 1)}(P,q,l)~=~ 2{\rm Re}\; \left[\, \sum_{i,j=1,2} \I^{P_iP_j}(P,q,l)\, \right]\, ,
\label{tot}
\eea
where the first superscript, $P_i$, on the right identifies the gluon
coupling in the
amplitude, to the quark,  $P_1$, or antiquark, $P_2$, and the
second, $P_j$, the gluon coupling in the complex conjugate.
As in Ref.\ \cite{fact2},
 we fix the 
 momentum of the eikonal $l$ in the minus light cone  direction,
 \bea
 l^\mu = \delta_{\mu -}\, .
 \eea
  The momenta $P_1$ and $P_2$ remain arbitrary.

  \subsubsection{The $P_1P_1$ gluon rescattering diagram}

 We begin with the diagram $\I^{P_1P_1}$, 
 Fig.\ \ref{fourdiagrams}a, which describes the 
 interference between the order $g_s^3$ rescattering of a gluon 
 that is emitted by the heavy quark ($P_1$) eikonal line
  and the lowest-order process in which it is emitted by the same line.

 Written out with the momentum structure of the eikonal  
vertices in Eq.~(\ref{eikrules}) shown explicitly, this
  diagram is given by
\bea
\I^{P_1P_1}(P,q,l)~
&=&
~\int \frac{d^D k_1}{(2\pi)^D} \frac{d^D k_2}{(2\pi)^D}~2\pi\, \delta(k_2^2)\,
\frac{-i}{[(k_2-k_1)^2 + i\epsilon]}~\frac{-i}{[k_1^2 + i\epsilon]}
\nonumber\\
\nonumber \\
&\ & \times\
\left( \, -g\mu^\vep\, V_{\nu,\mu,\lambda}[k_1,k_2-k_1,-k_2]\, \right)  
\nonumber \\
\nonumber \\
&\ &  \times\
\frac{i(-ig\mu^\vep)\, P_1^\nu}{[P_1 \cdot k_1 + i \epsilon]}~
\frac{-i( ig\mu^\vep)\, P_1^\lambda}{[P_1 \cdot k_2 - i\epsilon]}~
\frac{-i(-g\mu^\vep\ell^\mu)}{[l \cdot (k_2-k_1) + i\epsilon]
} \, ,
\label{sl1}
\eea
where the three-gluon vertex (with all momenta flowing in) 
is 
\be
V_{\mu_1,\mu_2,\mu_3}[q_1,q_2,q_3]~=~\, (q_1-q_2)_{\mu_3}~g_{\mu_1\mu_2}+
(q_2-q_3)_{\mu_1}~g_{\mu_2\mu_3}+
(q_3-q_1)_{\mu_2}~g_{\mu_3\mu_1}\, .
\ee
As in Eq.\ (\ref{P1P2Pq}), $P$ and $q$ are the total and relative 
momenta of the 
pair.  As in $\I^{(8\to 1)}$, Eq.\ (\ref{Edef}), 
$P_1$ and $P_2$ are the momenta of the heavy quark
and antiquark, respectively, and $l$
is the momentum of the lightlike eikonal.
As indicated in the diagram, we take
$k_1$ as the momentum of the soft gluon emitted in the amplitude,
to the left of the cut in Fig.\ \ref{fourdiagrams},
 and  $k_2$ as the momentum of the
soft gluon flowing to the right in the figure.
\footnote{In the discussion that follows, we will compute 
diagrams with loops in the amplitude, in contrast to the
complex conjugate amplitude as in Ref.\ \cite{fact2}.
Since our result is real,  the analysis is otherwise
completely equivalent.}

Following the procedure of
Ref.\ \cite{fact2}, we find it useful to integrate first the minus components
of $k_1$ and $k_2$.
Performing the $k_2^-$ integration by using the mass-shell delta function, 
and evaluating the numerator factors,  we find
\bea
&&\I^{P_1P_1}(P,q,l) = i~\frac{g^4\mu^{4\vep}}{(2\pi)^{2D-1}} 
~\int_{-\infty}^\infty d^{D-2} k_1^T \int_{-\infty}^\infty dk_1^+ \int_{-\infty}^\infty dk_1^- 
\int_{k_2^T<\mu} d^{D-2} k_2^T~ 
\int_0^\mu \frac{dk_2^+}{2k_2^+}~ \nonumber \\
&& \hspace{3mm}
\times \left [\, \left (P_1^+(\frac{{k_2^T}^2}{2k_2^+}+k_1^-)+P_1^-(k_2^+
+k_1^+)-P_1^T \cdot (k_2^T+k_1^T)\, \right ) 
-\frac{P_1^2}{P_1^+}(k_2^++k_1^+)\, \right ]  
~\frac{1}{[k_2^+ - k_1^+ + i\epsilon]}~\nonumber \\
&& \hspace{3mm}
\times 
\frac{1}{\left[
\frac{P_1^+ {k_2^T}^2}{2k_2^+}+P_1^-k_2^+-P_1^T \cdot k_2^T -i \epsilon
\right ]}
~\frac{1}{\left [2(k_1^+-k_2^+)(k_1^-- \frac{{k_2^T}^2}{2k_2^+}) 
~- {(k_1^T-k_2^T)}^2+i\epsilon\right ]} ~
\nonumber \\ 
&& \hspace{3mm}
\times 
\frac{1}{[2k_1^+k_1^-- {k_1^T}^2 +i\epsilon]} \;
\frac{1}{\left[
k_1^-+\frac{P_1^-k_1^+}{P_1^+}-\frac{P_1^T \cdot k_1^T}{P_1^+} +i \epsilon
\right ]}\, .
\label{sl3}
\eea
Below we will generally suppress limits on the loop ($k_1$) and 
real-gluon momentum  ($k_2$) integrals, except where they are necessary for the argument.
The $k_1$ integrals are unbounded, while
 the $k_2$ integrals are understood to be cut off at the order of the  quark mass.  
These upper limits 
are set by the phase space cutoff, 
denoted $\mu$ here, and
discussed above in connection with the 
basic matrix element of Eq.~(\ref{calM}).

The next step is the loop integral  $k_1^-$ in Eq.\ (\ref{sl3}), which 
we perform by contour integration.
The integrand of (\ref{sl3}) has three poles in $k_1^-$,
\bea
k_1^-{}_{[k_1^2]} &=&  \frac{k_{1}^T{}^2~-~i \epsilon}{2k_1^+}  \, ,
\nonumber\\
k_1^-{}_{[(k_1-k_2)^2]} &=&
 \frac{k_{2}^T{}^2}{2k_2^+} 
+ \frac{(k_1^T-k_2^T)^2~-~i\epsilon }{2(k_1^+-k_2^+)} \, ,
\nonumber\\
k_1^-{}_{[P_1\cdot k_1]} &=& 
\frac{P_1^T \cdot k_1^T}{P_1^+} -\frac{P_1^-k_1^+}{P_1^+}~-~i\epsilon \, ,
\label{kminuspoles}
\eea
labeled according to the denominator that vanishes.  

The pattern of poles
encountered here is similar to those for the order-$v^2$ calculation of
Ref.\ \cite{fact2},  and the role of each term is similar. 
When we close  the $k_1^-$ contour in the lower half-plane,
we pick up  the $k_1^-{}_{[k_1^2]}$ pole when  $k_1^+>0$,
the $k_1^-{}_{[(k_1-k_2)^2]}$ pole when $k_1^+>k_2^+$ and
the $k_1^-{}_{[p_1\cdot k_1]}$ pole for all values of $k_1^+$.
The contribution from the $k_1^-{}_{[k_1^2]}$ pole
vanishes, because the resulting integral is antisymmetric under the 
exchange of the remaining components of $k_1$  and $k_2$.
The $k_1^-{}_{[(k_1-k_2)^2]}$
contribution, 
on the other hand, cancels  against
the  corresponding cut in which the gluon with
momentum $k_1-k_2$ appears in the final state, as in Fig. 2b.

This leaves us with the third, $k_1^-{}_{[P_1\cdot k_1]}$, pole only,
whose calculation we describe in detail below for $\I^{P_1P_1}(P,q,l)$
and $\I^{P_1P_2}(P,q,l)$. We note that in the sum of
$\I^{P_1P_2}(P,q,l)$ and $\I^{P_2P_1}(P,q,l)$, the
symmetry argument for the cancellation of the $k_1^-{}_{[k_1^2]}$ pole
continues to apply.

In summary, to derive the infrared
contribution to $\I^{P_1P_1}(P,q,l)$ from Fig.\ \ref{fourdiagrams}, 
we close the 
$k_1^-$ integration in
Eq.\ (\ref{sl3}) in the 
lower  
half-plane at the pole $k_1^-{}_{[P_1\cdot k_1]}$,
and find
\bea
&&\I^{P_1P_1}(P,q,l) ~=~ (-2\pi i)\, (i)\,  
\frac{g^4\mu^{4\varepsilon}}{(2\pi)^{2D-1}} 
~\int d^{D-2} k_1^T \int dk_1^+ \int d^{D-2} k_2^T~ 
\int \frac{dk_2^+}{2k_2^+}~ \nonumber \\
&& \hspace{3mm}
\times
\left[ P_1^+\left (\frac{{k_2^T}^2}{2k_2^+}+
\left (\frac{P_1^T \cdot k_1^T}{P_1^+} -\frac{P_1^-k_1^+}{P_1^+}\right) \right)
+P_1^-(k_2^+
+k_1^+)-P_1^T \cdot (k_2^T+k_1^T)-\frac{P_1^2}{P_1^+}(k_2^++k_1^+) \right ]  
\nonumber \\
&& \hspace{3mm}
\times\
\frac{-1}{[k_1^+-k_2^+-i\epsilon]}~
~\frac{1}{\left [
\frac{P_1^+ {k_2^T}^2}{2k_2^+}+P_1^-k_2^+-P_1^T \cdot k_2^T -i \epsilon
\right ]} 
\nonumber \\
&& \hspace{8mm}
\times\ \frac{1}{\left [
2(k_1^+-k_2^+)\left ( \left (\frac{P_1^T \cdot k_1^T}{P_1^+} -\frac{P_1^-k_1^+}{P_1^+} \right)
- \frac{{k_2^T}^2}{2k_2^+} \right) ~- {\left (k_1^T-k_2^T\right )}^2+i\epsilon \right ]} 
~\nonumber \\
~&& \hspace{8mm}
\times\
\frac{1}{\left [
2k_1^+ \left (\frac{P_1^T \cdot k_1^T}{P_1^+} -\frac{P_1^-k_1^+}{P_1^+} \right)
- {k_1^T}^2 +i\epsilon \right ]} \, .
\label{sl4}
\eea
In this expression we have two ($D$-2 dimensional) transverse and two 
plus integrals remaining.
We begin by applying a  Feynman parametrization to the final two denominators,
\bea
&&\I^{P_1P_1}(P,q,l) ~=~ - \frac{g^4\mu^{4\varepsilon}}{(2\pi)^{2D-2}} 
~\int_0^1 dx 
\int d^{D-2} k_2^T \int dk_1^+ ~ 
\int \frac{dk_2^+}{2k_2^+}~\int d^{D-2} k_1'{}^T \nonumber \\
&& \hspace{15mm}
\times\
\Bigg [ \, P_1^+ \left (\frac{{k_2^T}^2}{2k_2^+}+
\left (\frac{P_1^T \cdot (k_1'{}^T+K^T(P_1))}{P_1^+} 
-\frac{P_1^-k_1^+}{P_1^+} \right) 
\right)+P_1^-(k_2^+ +k_1^+)
\nonumber\\
&& \hspace{35mm}
-P_1^T \cdot (k_2^T+k_1'{}^T+K^T(P_1)) 
-\frac{P_1^2}{P_1^+}(k_2^++k_1^+) \Bigg ]  \nonumber \\
&& \hspace{15mm}
\times\
\frac{1}{[k_1^+-k_2^+-i\epsilon]}~
~\frac{1}{ \left [\frac{P_1^+ {k_2^T}^2}{2k_2^+}+P_1^-k_2^+-P_1^T \cdot k_2^T 
-i \epsilon \right]} \nonumber \\
&& \hspace{15mm}
\times\
\frac{1}{[{k_1'{}^T}^2+L(P_1)-i\epsilon]^2}\, ,
\label{sl5}
\eea
where to complete the square we have shifted to $k_1'{}^T=k_1^T - K^T(P_1)$, 
with
\bea
K^T(P_1) \equiv 
x\, \left (k_2^T+\frac{(k_1^+-k_2^+)}{P_1^+}\, P_1^T \right )
+(1-x)\frac{k_1^+}{P_1^+}\,P_1^T \, .
\eea
In adddition, in the final denominator of (\ref{sl5}) we introduce the function
\bea
L(P_1)~\equiv~x\, \left [ \frac{k_1^+}{k_2^+}-x \right ]
\left [ \left (k_2^T
-\frac{k_2^+}{P_1^+}\,P_1^T  \right )^2 
+ \frac{k_2^+ k_1^+ }{x{P_1^+}^2}\, P_1^2 \right ] \, .
\label{L3def}
\eea
Here and below, to simplify the notation we suppress the $k_1$ and $k_2$ dependence
in $L(P_1)$ and $K^T(P_1)$.

We now perform the $k_1'{}^T$ integration 
of Eq.\ (\ref{sl5}) in $D$=4-2$\vep$ dimensions, obtaining
\bea
&&\I^{P_1P_1}(P,q,l)~ =~-\, \pi^{1-\varepsilon}\, \Gamma(1+\varepsilon)
\; \frac{g^4\mu^{4\varepsilon}}{(2\pi)^{2D-2}} 
~\int_0^1 dx \int dk_1^+ \int d^{D-2} k_2^T~ 
\int \frac{dk_2^+}{2k_2^+}~ \nonumber \\
&& \hspace{35mm}
\times
\left [ \left( P_1^+\frac{{k_2^T}^2}{2k_2^+}+ P_1^-k_2^+
-P_1^T \cdot k_2^T \right) -\frac{P_1^2}{P_1^+}(k_2^++k_1^+) \right ]  
\nonumber \\
&& \hspace{35mm}
\times
\frac{1}{[k_1^+-k_2^+-i\epsilon]}~
~\frac{1}{ \left [\frac{P_1^+ {k_2^T}^2}{2k_2^+}
                  +P_1^-k_2^+-P_1^T \cdot k_2^T -i \epsilon \right ]} 
\nonumber \\
&&\hspace{35mm}
\times \frac{1}{[L(P_1)-i\epsilon]^{1+\varepsilon}}\, .
\label{sl6}
\eea
Simplifying this expression algebraically, we put it into a form that will
facilitate the combination of diagrams below,
\bea
&&\I^{P_1P_1}(P,q,l)~ =~
-\, \pi^{1-\varepsilon}\, \Gamma(1+\varepsilon)
\; \frac{g^4\mu^{4\varepsilon}}{(2\pi)^{2D-2}} 
~\int_0^1 dx \int d^{D-2} k_2^T~ 
\int \frac{dk_2^+}{2k_2^+}~\int dk_1^+ 
\nonumber \\
&&  \hspace{15mm}
\times\frac{1}{[k_1^+-k_2^+-i\epsilon]}~ 
\frac{1}{[L(P_1)-i\epsilon]^{1+\varepsilon}} \;
\left [1  -\frac{\frac{P_1^2}{P_1^+}(k_2^++k_1^+)}
                {\frac{P_1^+ {k_2^T}^2}{2k_2^+}
+P_1^-k_2^+-P_1^T \cdot k_2^T -i \epsilon } \right ]
\, .
\label{sl7}
\eea
Of particular interest in this expression is the first term, 1, 
in square brackets.
This term has up to four poles in dimensional regularization, corresponding
to momentum configurations in which both momenta 
$k_1$ and $k_2$ vanish and are collinear to the light-like eikonal line $l$
in Fig.\ \ref{fourdiagrams}.  We will see that, as in Ref.\ \cite{fact2}, these
singularities cancel, leaving only a single real $1/\varepsilon$ (and
imaginary $1/\varepsilon^2$) pole in dimensional regularization.

\subsubsection{The $P_1P_2$ diagram and collinear cancellation}

Before continuing with the integrals, we turn our attention to the 
third diagram in Fig.\ \ref{fourdiagrams}a,
which describes interference between gluon ($k_1$) rescatterring 
after emission from the heavy quark ($P_1$) line in the
amplitude with
 gluon ($k_2$) emission from the antiquark ($P_2$) line in the complex conjugate
amplitude.   We denote this diagram by $\I^{P_1P_2}(P,q,l)$.

The momentum-space integral for the $P_1P_2$
rescattering diagram, $\I^{P_1P_2}(P,q,l)$ is
\bea
\I^{P_1P_2}(P,q,l)~
&=&   -i\, g^4 \mu^{4\varepsilon}\;
\int \frac{d^D k_2}{(2\pi)^D} \frac{d^D k_1}{(2\pi)^D}~2\pi~\delta(k_2^2)
  \nonumber \\
&\ & \hspace{-5mm} 
\times P_1^\nu l^\mu P_2^\lambda \, V_{\nu,\mu,\lambda} [k_1,k_2-k_1,-k_2] 
\nonumber\\
&\ & \hspace{-5mm}
\times
\frac{1}{[P_1 \cdot k_1 + i \epsilon]~[P_2 \cdot k_2 -i\epsilon]
~[l \cdot (k_2-k_1) +i\epsilon]~
[(k_1-k_2)^2 + i\epsilon]~[k_1^2 + i\epsilon]} 
\, .
\nonumber \\
\label{sl1a}
\eea
This diagram has an overall $(-1)$ relative to $\I^{P_1P_1}(P,q,l)$, associated
with the connection of one gluon to the $P_2$ eikonal line, which is
in the antiquark representation.

Performing on $\I^{P_1P_2}(P,q,l)$
the same steps as above for the $k_i^-$ and $k_1^T$ 
integrals 
in $\I^{P_1P_1}(P,q,l)$,
we can put this integral into a form analogous to Eq.\ (\ref{sl7}),
although slightly more complex,
\bea
&&\I^{P_1P_2}(P,q,l)  =~  \pi^{1-\varepsilon} \Gamma(1+\varepsilon)
\; \frac{g^4\mu^{4\varepsilon}}{(2\pi)^{2D-2}}  
~\int_0^1 dx  \int d^{D-2} k_2^T~ \int \frac{dk_2^+}{2k_2^+}~\int dk_1^+
\nonumber\\
&& \hspace{10mm} 
\times \frac{1}{[k_1^+-k_2^+-i\epsilon]}
\frac{1}{[L(P_1)-i\epsilon]^{1+\varepsilon}} \ \bigg [\ 1 \ -
\nonumber \\
\ 
\nonumber\\
&&
\frac{ 
\frac{P_1 \cdot P_2}{P_1^+}(k_1^++k_2^+)
+\frac{2}{P_1^+}(k_1^+-k_2^+)(P_2^+P_1^--P_1^+P_2^-)
+2\, P_2^+\, \left[ \frac{P_2^T}{P_2^+}-\frac{P_1^T}{P_1^+}\right]\cdot 
             \left[K^T(P_1)-k_2^T \right]
}
{\left [\frac{P_2^+ {k_2^T}^2}{2k_2^+}
+P_2^-k_2^+-P_2^T \cdot k_2^T -i \epsilon\right ]} \bigg ].
\nonumber\\
\label{sl8}
\eea
Combining the  expressions for
$\I^{P_1P_1}(P,q,l)$ and $\I^{P_1P_2}(P,q,l)$, 
Eqs.\ (\ref{sl7}) and (\ref{sl8}),
we see that,
as anticipated above, the collinear-singular terms (the 1s) 
in the square brackets cancel. 

Our goal now is to evaluate the infrared poles
from the expressions in Eqs.~(\ref{sl7}) and (\ref{sl8}). 
For this purpose we need to perform the $k_1^+$ integration,
which ranges from -$\infty$ to +$\infty$. 
Noting that the numerators are at most linear in $k_1^+$,
we rewrite factors of $k_1^+$ in the numerator as
$(k_1^+-k_2^+) +k_2^+$. The $(k_1^+-k_2^+)$ term then
cancels the corresponding denominator in Eq.\ (\ref{sl8}).
We  next
reorganize
the contributions of the $P_1P_1$ and $P_1P_2$
diagrams to $\I^{(8\to 1)}$ as
\bea
 {\cal I}^{P_1P_2}\ + \ {\cal I}^{P_1P_1} 
 =({\cal J}^{P_1P_2}\ + \ {\cal J}^{P_1P_1}) +
 ({\cal K}^{P_1P_2}\ + \ {\cal K}^{P_1P_1}) \, ,
\label{jk}
\eea
where, for example,
 ${\cal J}^{P_1P_2}$ is obtained from  ${\cal I}^{P_1P_2}$ by canceling the ($k_1^+-k_2^+$) 
factors in the numerator
and denominator, and ${\cal K}^{P_1P_2}$ 
represents the remaining terms in ${\cal I}^{P_1P_2}$. 
We now turn to the identification of infrared
poles in these expressions, using slightly different
procedures in the two cases.

\subsection{The IR pole from the ${\cal J}$ terms}

We begin with the terms that lack the pole in $k_1^+-k_2^+$.
From Eq.~(\ref{sl8}) we find
\bea
&& \hspace{-10mm} {\cal J}^{P_1P_2}(P,q,l)   
+{\cal J}^{P_1P_1}(P,q,l)   
= ~ - \pi^{1-\varepsilon} \Gamma(1+\varepsilon)
\; \frac{g^4\mu^{4\varepsilon}}{(2\pi)^{2D-2}} 
~\int_0^1 dx \int dk_2^+ \int d^{D-2} k_2^T
\nonumber\\
&& \hspace{35mm} \times  \int_{-\infty}^\infty dk_1^+ \
\frac{ 
\frac{P_1 \cdot P_2}{P_1^+P_2^+}
+\frac{2}{P_1^+P_2^+}(P_2^+P_1^--P_1^+P_2^-)
+2\, \left[ \frac{P_2^T}{P_2^+}-\frac{P_1^T}{P_1^+}\right]\cdot 
             \frac{P_1^T}{P_1^+} 
}{
{[L(P_1)-i\epsilon]^{1+\varepsilon}} \ 
\left [{k_2^T}^2
+\frac{P_2^-}{P_2^+}2{k_2^+}^2-\frac{P_2^T}{P_2^+}2k_2^+ \cdot k_2^T -i \epsilon\right ]}
\nonumber\\
&& \hspace{50mm} -(P_2 \rightarrow P_1)\, .
\label{sla8}
\eea
In these terms, the remaining (quadratic) $k_1^+$-dependence
is in the $L(P_1)$ denominator.  The integral is elementary, and
we find
\bea
&&{\cal J}^{P_1P_2}(P,q,l)   
+{\cal J}^{P_1P_1}(P,q,l)   
=~ - 2^{1+2\vep}(-1-i\epsilon)^{-1/2-\vep}~B(1/2,1/2+\vep)~ \pi^{1-\varepsilon} \Gamma(1+\varepsilon)
\;  \nonumber \\
&& \hspace{45mm} \times\ \frac{g^4\mu^{4\varepsilon}}{(2\pi)^{2D-2}}  (\frac{P_1^2}{{P_1^+}^2})^{\vep}
~\int_0^1 \frac{dx}{x^{1+2\vep}} 
\int_0^\mu \frac{dk_2^+ }{{k_2^+}^{1+2\vep}}~
\int_{k_2^T<\mu} \frac{d^{D-2} k_2^T~ }{{k_2^+}^2}
\nonumber\\
&& \hspace{45mm} \times 
\frac{ 
\frac{P_1 \cdot P_2}{P_1^+P_2^+}
+\frac{2}{P_1^+P_2^+}(P_2^+P_1^--P_1^+P_2^-)
+2\, \left[ \frac{P_2^T}{P_2^+}-\frac{P_1^T}{P_1^+}\right]\cdot 
             \frac{P_1^T}{P_1^+} 
}{
\left [(\frac{k_2^T}{k_2^+}-\frac{P_1^T}{P_1^+})^2+\frac{P_1^2}{{P_1^+}^2} \right]^{1+2\vep}
\left [(\frac{k_2^T}{k_2^+}-\frac{P_2^T}{P_2^+})^2+\frac{P_2^2}{{P_2^+}^2} \right]
} \nonumber \\
&& \hspace{45mm} -(P_2 \rightarrow P_1)\, .
\label{sla9}
\eea
The leading behavior of this expression is a purely imaginary double
pole in $\vep$, from the lower limits of the $x$ and $k_2^+$ 
integrals.  The remaining transverse integration is both infrared
and ultraviolet finite, and real, for $\vep \to 0$, so that an expansion
in $\vep$ of this integral can 
give rise to
only imaginary single poles,
which vanish in the cross section.  The overall factor of $(-1-i\epsilon)^{-\vep}$,
however, can 
convert an imaginary double pole to 
a {\it real}, single infrared pole.  To isolate the residue of this
pole, we need only evaluate the 
$k_2^T$ 
transverse integral at $\vep=0$.
An important point is that, because we are interested only in noncancelling
infrared poles, we may extend the upper limit of the $k_2^T$ integral to
infinity.  We can do this because at finite $k_2^T>\mu$, only the
$k_1$ line can produce an infrared pole, at one loop.  But, as discussed
in \cite{fact2}, for example, all one-loop infrared divergences
factorize in the sense of NRQCD.

A simple change of variables, $y_T=k_2^T/k_2^+$, simplifies the transverse integration,
evaluated at $\vep=0$ as just described,
\bea
&& \int  \frac{d^2k_2^T}{\left(k_2^+\right)^2}\; \frac{1}{\left [
\left (\frac{k_2^T}{k_2^+}-\frac{P_1^T}{P_1^+} \right)^2+\frac{P_1^2}{{P_1^+}^2} \right]
\left [ \left (\frac{k_2^T}{k_2^+}-\frac{P_2^T}{P_2^+} \right)^2+\frac{P_2^2}{{P_2^+}^2} \right] } 
\nonumber\\
&& \hspace{40mm} =
\  \left(\, \frac{{P_1^+}^2}{P_1^2}~\, \right)\; \int d^2y_T ~ 
\frac{1}{\left [y_T^2+1 \right]
\left [(y_T+a_T)^2+\frac{P_2^2}{{P_2^+}^2} \frac{{P_1^+}^2}{{P_1}^2} \right] } \, ,
\eea
where we define
\bea
a_T \equiv \left ( \frac{P_1^T}{P_1^+}-\frac{P_2^T}{P_2^+} \right) 
\sqrt{\frac{{P_1^+}^2}{{P_1}^2}} \, .
\eea
Next, introducing another Feynman parametrization, $w$ for $y_T$
and integrating over $y_T$ we get
\bea
&& \int d^2y_T ~ 
\frac{1}{\left [y_T^2+1 \right]
\left [(y_T+a_T)^2+\frac{P_2^2}{{P_2^+}^2} \frac{{P_1^+}^2}{{P_1}^2} \right] } 
 \nonumber \\
&& \hspace{40mm}
= \pi\ \int_0^1 dw \frac{1}{\left [1-w-w^2a_T^2+w(a_T^2+
\frac{P_2^2}{{P_2^+}^2} \frac{{P_1^+}^2}{{P_1}^2}) \right] } \, .
\eea
After integrating over $w$ we then find, for the original $k_2^T$ integral,
\bea
&& \int \frac{d^2k_2^T}{\left(k_2^+\right)^2}~ \frac{1}{\left [
\left (\frac{k_2^T}{k_2^+}-\frac{P_1^T}{P_1^+} \right )^2+\frac{P_1^2}{{P_1^+}^2} \right]
\left [ \left (\frac{k_2^T}{k_2^+}-\frac{P_2^T}{P_2^+} \right)^2+\frac{P_2^2}{{P_2^+}^2} \right] }
~=~\frac{\pi}{c}\ln\left [\frac{a+c}{a-c} \right ]\, ,
\label{in11}
\eea
where
\bea
&& a~\equiv~\frac{2P_1 \cdot P_2}{P_1^+P_2^+} \nonumber \\
&& c~\equiv~a ~\sqrt{ 1~ - ~ \frac{P_1^2 P_2^2}{(P_1 \cdot P_2)^2}} \, .
\label{in22}
\eea
We are now ready to isolate the real single pole, as indicated above.
The poles of the remaining $x$ and $k_2^+$ integrations in Eq.\ (\ref{sla9}) are found
by using $1/{f^{1+a}} =-{\delta(f)}/{a}$,
and for the $\vep$-dependent phase, the expansion
\bea
(-1 \pm i\epsilon)^{\varepsilon}=e^{\pm i\pi \varepsilon}=1 \pm i\pi \varepsilon + {\cal O} (\varepsilon^2)+
\dots\, 
\label{exp1}
\eea
gives the corresponding $\vep$ term.
We then find from Eq. (\ref{sla9}), adding the two remaining diagrams (which are
determined by simple substitution),
\bea
&& 2\,{\rm Re}
 \left[{\cal J}^{P_1P_2}\ + \ {\cal J}^{P_2P_1} 
      +{\cal J}^{P_1P_1}\ + \ {\cal J}^{P_2P_2} \right] \nonumber \\
&& \hspace{45mm} 
= \frac{\alpha_s^2}{4 \varepsilon}\,
 \left [\,\left(1+\frac{P_1^+ P_2^+}{P_1 \cdot P_2} \Delta
  P^2\right)\; 
\frac{1}{2h}\, \ln \left (\frac{1+h}{1-h} \right ) -1 \right]\, ,
\label{pdpj}
\eea
where $\Delta P^2$ and $h$ depend on the momenta of the pair as
\bea
&& 
\Delta P^2~\equiv~\left(\frac{P_1}{P_1^+}-\frac{P_2}{P_2^+}\right)^2  
\, ,
\nonumber \\
&&
h~\equiv ~\sqrt{ 1~ - ~ \frac{P_1^2 P_2^2}{(P_1 \cdot P_2)^2}} \, .
\label{hh}
\eea
It is easy to see that $h$ is proportional to the velocity of
the heavy quarks in the center of mass system.  We will 
give an explicit expression
for its full $v$-dependence
 below, after identifying the poles of the
$\cal K$ terms.
Taken in isolation, however, the sum of the $\cal  J$ terms is
manifestly not independent of the choice of $l^\mu$, because of the
explicit dependence on the factors $P_i^+ = P_i\cdot \ell$ in (\ref{pdpj}). 
\footnote{
It is interesting to observe that the overall result in (\ref{pdpj})  is invariant under independent
rescalings of $P_1$, $P_2$ and  $l$, as expected for the eikonal approximation.}
Independence from the direction of the light-like 
Wilson line used to define the underlying matrix element will, however, emerge 
below, once we find the
infrared poles of the $\cal K$ terms.

\subsection{The IR pole from the ${\cal K}$ terms}

Now we evaluate the remaining terms in Eq.\ (\ref{sl8}), 
for which
the factor $k_1^+-k_2^+$ remains in the
denominator, such as
${\cal K}^{P_1P_2}\ = \ {\cal I}^{P_1P_2} - \ {\cal J}^{P_1P_2} $.  
In these cases, the virtual $k_1^+$ dependence in the denominators
is no longer quadratic, and the integral requires a slightly more elaborate approach.
Specifically, the contributions from the $P_1P_2$ and $P_1P_1$
diagrams are given by
\bea
&&{\cal K}^{P_1P_2}(P,q,l)  
+{\cal K}^{P_1P_1}(P,q,l)  
=~ - \pi^{1-\varepsilon} \Gamma(1+\varepsilon)
\; \frac{g^4\mu^{4\varepsilon}}{(2\pi)^{2D-2}}  \nonumber \\
\ \nonumber \\
&& \hspace{40mm} \times\ \int_0^1dx ~\int \frac{dk_2^+}{2k_2^+}~  \int dk_1^+ \int d^{D-2} k_2^T~ 
 \frac{1}{[k_1^+-k_2^+-i\epsilon]}
\frac{1}{[L(P_1)-i\epsilon]^{1+\varepsilon}} \nonumber \\
\ \nonumber \\
&& \hspace{40mm} \times\ \bigg [ \frac{ \frac{P_1 \cdot P_2}{P_1^+}(2k_2^+)
+2\, P_2^+\, \left[ \frac{P_2^T}{P_2^+}-\frac{P_1^T}{P_1^+}\right]\cdot 
             \left[G^T(P_1)-k_2^T \right]
}
{\left [\frac{P_2^+ {k_2^T}^2}{2k_2^+}
+P_2^-k_2^+-P_2^T \cdot k_2^T -i \epsilon\right ]} \bigg ]
\nonumber\\
\ \nonumber \\
&& \hspace{50mm} 
-\ (P_2 \rightarrow P_1)\, .
\label{sl8r}
\eea
where
\bea
G^T(P_1) \equiv
x\, k_2^T+(1-x)k_2^+\frac{P_1^T}{P_1^+} \, .
\eea
At this stage we employ another Feynman parametrization 
($y$) to organize and perform the $k_2^T$ integration.  As for the $\cal J$ terms,
we may extend the upper limit of the transverse integral to infinity,
 with the result
\bea
&& {\cal K}^{P_1P_2}\ + \ {\cal K}^{P_1P_1} 
\nonumber\\
&& \hspace{20mm} =
\frac{-1}{32 \pi^4} ~
 g^4(4\pi\mu^2)^{2\varepsilon}\,\Gamma(1+2\varepsilon)~
\int_0^1 dx \int_0^1 dy y^\varepsilon 
\int dk_1^+ ~ \int \frac{dk_2^+}{k_2^+}
\frac{k_2^+}{[k_1^+-k_2^+-i\epsilon]}~ \nonumber \\
\ \nonumber \\
&& \hspace{35mm} \times\ \Bigg [k_2^+ \left( \frac{P_1\cdot P_2}{P_1^+P_2^+} +
\Delta P^2(1-x)(1-y)\right) \Bigg ]\;
\frac{1}{{(M-i\epsilon)}^{1+\varepsilon} 
{(N+i\epsilon)}^{1+2\varepsilon}}
\nonumber \\ 
\ \nonumber \\
&& \hspace{25mm} \quad - \
(P_2 \rightarrow P_1 ) \, ,
\label{sl9}
\eea
where we define
\bea
&& M~\equiv~x \left (\frac{k_1^+}{k_2^+}~-~x \right) \, ,
 \nonumber \\
&& N~\equiv~{k_2^+}^2
\left [-y(1-y) \Delta P^2 +\frac{y}{x} 
\left(\frac{k_1^+ P_1^2}{k_2^+ {P_1^+}^2} - (x-x/y)
\frac{{P_2^2}}{{P_2^+}^2} \right)\, \right ]\, ,
\eea
with $\Delta P^2$ defined in Eq.~(\ref{hh}).
In Eqs.\ (\ref{sl8r}) and (\ref{sl9}), we have exhibited  ${\cal K}^{P_1P_2}$, from the third figure in the diagram
Fig 2a. The other term may be found as indicated, simply by
replacing $P_2$ everywhere by $P_1$, which leads to a
number of simplifications. For example, $\Delta P^2$ is replaced by zero
in these terms.

Equation (\ref{sl9}) can be reorganized slightly to yield
\bea
&&{\cal K}^{P_1P_2}\ + \ {\cal K}^{P_1P_1} 
\nonumber\\
&& \hspace{20mm} =
\frac{-1}{32 \pi^4} ~
g^4(4\pi\mu^2)^{2\varepsilon}\, \Gamma(1+2\epsilon)~
~\int_0^1 dx \int_0^1 dy y^\varepsilon ~ 
\int_0^\mu \frac{dk_2^+}{{k_2^+}^{1+4\varepsilon}}~ 
\int_{-\infty}^\infty \frac{dz}{z-1-i\epsilon}
 \nonumber \\
 \ \nonumber\\
&& \hspace{20mm} \times\ 
\frac{1}{{(x(z-x)-i\epsilon)}^{1+\varepsilon}~(y/x)^{1+2\varepsilon}}
\nonumber\\
 \ \nonumber\\
 && \hspace{20mm} \times\
\frac{1}{\left [-x(1-y)\Delta P^2 \frac{{P_1^+}^2}{P_1^2}
+x(1/y-1) \frac{{P_1^+}^2}{{P_2^+}^2} \frac{{P_2}^2}{P_1^2}+z 
  +i\epsilon \right]^{1+2\varepsilon}} \nonumber \\ 
  \ \nonumber\\
&& \hspace{20mm} \times\
\left(\frac{{P_1^+}^2}{P_1^2}\right)^{1+2\vep}\; \Bigg[
\left (\frac{P_1 \cdot P_2}{P_1^+P_2^+}
+\Delta P^2 (1-x)(1-y) \right ) \Bigg ] \nonumber \\
\ \nonumber \\
&& \hspace{20mm} 
\quad 
~-~(P_2 \rightarrow P_1 )\, ,
\label{sl10}
\eea
where we define $z=k_1^+/k_2^+$.
The variable $z$ appears in three denominators, one of which
is a simple pole at $z=1$ in the upper half plane.  
To perform the $z$ integral in Eq.\ (\ref{sl10}), we combine 
the remaining two denominators by introducing another
Feynman parameter, $y^\prime$,
\bea
&&\frac{1}{{(x(z-x)-i\epsilon)}^{1+\varepsilon}}\, 
\frac{1}{ \left [\, -x(1-y)\Delta P^2 \frac{{P_1^+}^2}{P_1^2}
+x(1/y-1) \frac{{P_1^+}^2}{{P_2^+}^2} \frac{{P_2}^2}{P_1^2}+z +i\epsilon
\, \right ]^{1+2\varepsilon}}
\nonumber\\
&& \ =  
 \frac{1}{x^{1+\varepsilon}}\; 
\frac{\Gamma(2+3\varepsilon)}
     {\Gamma(1+\varepsilon)\, \Gamma(1+2\varepsilon)} \nonumber \\
&&~ \times \int_0^1 dy'\ 
\frac{y'{\, }^\varepsilon\, (1-y')^{2\varepsilon}
      (-1+i \epsilon)^{-1 -2\varepsilon}}
     { \left[ (2y^\prime -1)z+x(1-y^\prime)
       \left((1-y) \Delta P^2 \frac{{P_1^+}^2}{P_1^2}-\frac{1-y}{y}
       \frac{{P_1^+}^2}{{P_2^+}^2} \frac{{P_2}^2}{{P_1}^2}\right)
       -xy^\prime -i\epsilon \right]^{2+3\varepsilon} }\, .
\label{yprimeparam}
\eea
Notice that before the combination of these two $z$-dependent denominators with
fractional powers, we factor $-1+i\epsilon$ from the
second of the two, so that the $i\epsilon$ on the
right-hand side of (\ref{yprimeparam}) has
a definite sign: $-i\epsilon$.    It is perhaps worth noting
further 
how we determine
this prescription.  As
$z$ varies  from -$\infty$ to +$\infty$, the second (and first) denominator 
in  (\ref{yprimeparam}) can vanish,
but for any variable denominator, denoted $W$, we have
\bea
&& ~W+i\epsilon = W  ~~~~~~~~~{\rm if} ~~~~~~~~~ W > 0   \nonumber \\
&& ~W+i\epsilon = (-1+i\epsilon)(-W-i\epsilon)  ~~~~~{\rm if} ~~~~~~W < 0.  ~~~~
\label{meps}
\eea
The factor $(-1+i\epsilon)^{\vep} = 1 +i\pi\vep+\dots$ in this expression 
will be particularly  important
in the calculation below, because
as above 
it will convert 
an imaginary double pole in $\vep$ to a real  single pole.

Although Eq.~(\ref{yprimeparam}) is a bit complicated overall, 
the combined denominator on the right-hand side
gives a branch cut in the $z$-plane that is in the lower (upper)
half-plane for $0<y'<1/2$ ($1/2<y'<1$).   
Inserting the expression in Eq.~(\ref{yprimeparam}) back into 
Eq.~(\ref{sl9}), we then easily 
perform the $z$ integral.  For $y'>1/2$ we
can complete the $z$ contour in the lower half plane without
enclosing any singularities, and the $z$ integral vanishes.
For $y'<1/2$ we close in the upper half plane and pick up
the simple pole at $z=1$.  

The result of this procedure is
\bea
&&{\cal K}^{P_1P_2}\ + \ {\cal K}^{P_1P_1}
\nonumber\\
&& \hspace{20mm} =~ 
\frac{-i(-1-i \epsilon)^{-1 -\varepsilon}}{16 \pi^3}~
\frac{\Gamma(2+3\vep)}{\Gamma(1+\vep)}
~ g^4(4\pi\mu^2)^{2\varepsilon}\,
\int \frac{dk_2^+}{{k_2^+}^{1+4\varepsilon}}
~\int_0^1 dx x^\varepsilon \int_0^1 dy y^{1+2\varepsilon}  \nonumber \\
&&\hspace{25mm} \times\ \int_0^{1/2} dy^\prime y^\prime{\, }^\varepsilon 
  (1-y^\prime)^{2\varepsilon} \times \ 
\left(\frac{{P_1^+}^2}{P_1^2}\right)^{1+2\vep}\; 
\left[\frac{P_1 \cdot P_2}{P_1^+P_2^+}
+\Delta P^2 (1-x)(1-y) \right ] \nonumber \\
&& \hspace{25mm} \times \
\frac{1}{ \left[\, (1-2y^\prime)y+x(1-y)(1-y^\prime) f(P_1,P_2) 
  +xyy^\prime +i\epsilon
  \, \right ]^{2+3\varepsilon}} \nonumber \\
\nonumber\\
&& \hspace{20mm} - ~~(P_2 \rightarrow P_1 )\, .
\label{sl11}
\eea
where
\bea
f(P_1, P_2)~=~
\frac{{P_1^+}^2}{{P_2^+}^2} \frac{{P_2}^2}{P_1^2} 
- y \Delta P^2 \frac{{P_1^+}^2}{P_1^2}\, .
\eea
In the above equation,
$(-1-i\epsilon)^{-1-\vep}$ is obtained by multiplying the overall factor 
$(-1+i\epsilon)^{-1-2\vep}$,  from Eq.\ (\ref{yprimeparam}), 
together with
$(-1-i\epsilon)^{-2-3\vep}$, from the denominator on the
right-hand side of (\ref{sl11}), after the $z$ integration. 
Although the result has the same
number of integrations, the infrared behavior of the
$y'$ integral is straightforward to analyze, which
will simplify our 
determination
of the infrared poles.

We can  check 
the infrared behavior of the $y^\prime$ 
integration by expanding
in $\varepsilon$,
\bea
y^\prime{\, }^\varepsilon (1-y^\prime)^{2\varepsilon} 
= 1+\varepsilon [\ln y^\prime + 2 \ln (1- y^\prime)]+ \dots \, .
\eea
The term in square brackets 
is finite at the endpoint $y^\prime = 1/2$.
Because the denominator 
is finite at $y^\prime =0$, the behavior 
at $y'=0$ is then no worse than logarithmic,
 $\int_0^{1/2} {dy^\prime \ln y^\prime }$, and 
the $y'$ integral is finite 
 and real
for $\varepsilon=0$.
 Thus, to identify the 
 real infrared pole, we can replace 
$y'{\, }^\varepsilon(1-y')^{2\varepsilon}$
by unity and integrate
over $y^\prime$ in (\ref{sl11}) to obtain
an expression that includes the full real single pole,
\bea
&&{\cal K}^{P_1P_2}\ + \ {\cal K}^{P_1P_1}
\nonumber\\
&& =~ i\, (-1-i \epsilon)^{-1 -\varepsilon}\, 
\frac{\Gamma(2+3\vep)}{(1+3\vep )~\Gamma(1+\vep)}
\frac{1}{16\pi^3}~ g^4(4\pi\mu^2)^{2\vep}\,
\int \frac{dk_2^+}{{k_2^+}^{1+4\varepsilon}}
~\int_0^1 dx x^\varepsilon \int_0^1 dy y^{1+2\varepsilon} ~ 
\nonumber \\
&&~ \times \ 
\frac{{P_1^+}^2}{P_1^2}\; \left (\, \frac{P_1 \cdot P_2}{P_1^+P_2^+}
+\Delta P^2 (1-x)(1-y) \right ) 
\times \ \frac{1}{[ -2y-x(1-y)f(P_1,P_2) +xy]} \nonumber \\
\nonumber\\
&& ~\left [\, 
\frac{1}{[\frac{1}{2}x(1-y) f(P_1,P_2) +\frac{1}{2}xy]^{1+3\varepsilon}}
-\frac{1}{[ y+x(1-y) f(P_1,P_2)]^{1+3\varepsilon}}\, 
\right ]\nonumber \\
\nonumber\\
&& -~~(P_2 \rightarrow P_1 )\, .
\label{sa11}
\eea
Here we note that terms of order $\vep$ from the $y$
integration can
also  produce at worst imaginary
single poles in $\vep$.   

The steps that isolate
the real infrared pole for the $\cal K$'s 
follow the procedure of the previous subsection.
The $x$ and $y$ integrals in Eq.\ (\ref{sa11}) integrals are real, and  
we must identify a $1/\varepsilon$ pole from these integrals.
The  resulting double pole then combines with
the $i\pi\varepsilon$ term from the expansion
of the prefactor $(-1-i \epsilon)^{-1 -\varepsilon}$
to give a real pole term that survives in 
the full sum of diagrams

It is easy to check that the 
$x$ and $y$ integrals of the
second fraction in  square brackets 
in Eq.~(\ref{sa11})
are both real and finite for $\varepsilon \to 0$.  We may thus
limit our analysis to the first fraction, which is proportional
to $1/x^{1+3\varepsilon}$.
The $x\to 0$ limit thus generates another infrared pole,
which can be isolated by using the distribution identity,
$1/{x^{1+a}} = -\delta(x)/a$,
while the $y$ integral can be performed at $\vep=0$.

For the  remaining analysis, it is convenient to combine 
Eq.\ (\ref{sa11}) with the corresponding results
from the 
other
two diagrams in
Fig.\ \ref{fourdiagrams}a, which 
can be found by  substitution.
After  performing the 
$x$  integration  in (\ref{sa11}) we obtain,  suppressing the finite part,
\bea
&&{\cal K}^{P_1P_2} + {\cal K}^{P_2P_1} 
+ {\cal K}^{P_1P_1} + {\cal K}^{P_2P_2}
\nonumber\\
&& \hspace{20mm}  = ~
\frac{i\, (-1-i \epsilon)^{-1 -\varepsilon}}{32 \pi^3 \vep}~
\frac{\Gamma(2+3\vep)}{(1+3\vep )~\Gamma(1+\vep)}
 2^{3\vep}~g^4~(4\pi\mu^2)^{2\varepsilon}\,
\int_0^\mu \frac{dk_2^+}{{k_2^+}^{1+4\varepsilon}} 
~\int_0^1 dy y^{2\varepsilon} ~ \nonumber \\
&&\hspace{25mm} \times\
 \left(\frac{{P_1^+}^2}{P_1^2}\right)^{1+2\vep}
 \left[\frac{P_1 \cdot P_2}{P_1^+P_2^+} +\Delta P^2 (1-y) \right]
\frac{1}{\left[y+(1-y) f(P_1,P_2) \right]^{1+3\varepsilon}} \nonumber \\
&& \hspace{25mm} +~~(P_1 \leftrightarrow P_2 ) 
~ - ~(P_2 \rightarrow P_1 ) ~ - ~(P_1 \rightarrow P_2 ) \, .
\label{sl12}
\eea
Taking into account the $k_2^+$ integral, we see that
this expression has an overall imaginary double pole which, however,
cancels when the complex conjugate diagrams are combined.

The leading, {\it real} pole in $\varepsilon$ can be
obtained from the above equation by setting $\vep \to 0$
in the $y$ integral, which then reduces to 
the sum of two elementary integrations.
The first has a $y$-independent numerator in the integrand,
\bea
&&~\int_0^1 dy \frac{{{P_1^+}^2}/{P_1^2}}{[y + (1-y)f(P_1,P_2)]}
~=~\int_0^1 dy \frac{{{P_2^+}^2}/{P_2^2}}{[y + (1-y)f(P_2,P_1)]} 
~=~\frac{1}{c}\ln\left [\frac{a+c}{a-c} \right ]\, ,
\label{int1}
\eea
with $a$ and $c$ defined in Eq.~(\ref{in22}).
The additional term is linear in $y$ in the numerator, and is given by
\bea
\int_0^1 dy ~y~\frac{{P_1^+}^2/{P_1^2}}{[y + (1-y)f(P_1,P_2)]}
~=~\frac{d(P_1,P_2)}{2c}\, \ln \left[\frac{a+c}{a-c} \right]~+~\frac{1}{2\Delta P^2} 
\ln \left [ \frac{P_1^2 {P_2^+}^2}{P_2^2 {P_1^+}^2}\right]\, ,
\label{int3}
\eea
where $a$ and $c$ are defined in (\ref{in22}), $\Delta P^2$ in (\ref{hh}) and
\bea
d(P_1,P_2)~\equiv~1-\frac{1}{\Delta P^2}(\frac{P_1^2}{{P_1^+}^2}~-~\frac{P_2^2}{{P_2^+}^2}).
\label{int4}
\eea
Although the integral of Eq. (\ref{int3}) is not symmetric in $P_1$ and $P_2$,
once we add the four diagrams we obtain symmetric result, equal to (\ref{int1}), 
\bea
&&~\int_0^1 dy ~y~\left [\frac{{{P_1^+}^2}/{P_1^2}}{ y + (1-y)f(P_1,P_2) }
~+~ \frac{{{P_2^+}^2}/{P_2^2}}{ y + (1-y)f(P_2,P_1) } \right ]
~=~\frac{1}{c}\, \ln \left[\frac{a+c}{a-c}\right]\, .
\label{int2b}
\eea

Combining the results above in Eq.\ (\ref{sl12}), 
and using
\bea
(-1 \pm i\epsilon)^{\varepsilon}=e^{\pm i\pi \varepsilon}=1 \pm i\pi \varepsilon + {\cal O} (\varepsilon^2)+
\dots ,
\label{expp}
\eea
we obtain for the infrared pole,
\bea
&& 2\,{\rm Re}\, \left[\,
  {\cal K}^{P_1P_2} + {\cal K}^{P_2P_1} 
+ {\cal K}^{P_1P_1} + {\cal K}^{P_2P_2}\, \right]
 =
 \frac{\alpha_s^2}{4 \varepsilon}~\left[\, 2 - 
       \left(\frac{2P_1 \cdot P_2}{P_1^+ P_2^+}+\Delta P^2\right)\; 
       \frac{1}{c}\, \ln \left (\frac{a+c}{a-c} \right ) \right]\, 
\nonumber\\
 && \hspace{55mm} =
 \frac{\alpha_s^2}{4 \varepsilon}~\left[\, 2 - 
       \left(\, 2 + \frac{P_1^+ P_2^+}{P_1 \cdot P_2} \Delta
       P^2\right)\; \frac{1}{2h}\, \ln \left (\frac{1+h}{1-h} \right ) 
       \right]\, ,
\label{pdpk}
\eea
with $h$ defined in Eq.~(\ref{hh}).
Again, we find a dependence on the choice of vector $l$, which,
however, cancels in the sum of the $\cal J$s and $\cal K$s.

\subsection{The IR pole in ${\cal I}^{8 \to 1}$}

Now adding Eqs. (\ref{pdpj}) and (\ref{pdpk}) we find from Eq.~(\ref{sl8})
\bea
\I^{(8\to 1)} 
&=&
2\,{\rm Re}\, \left[\,
  {\cal I}^{P_1P_2} + {\cal I}^{P_2P_1} 
+ {\cal I}^{P_1P_1} + {\cal I}^{P_2P_2}\, \right]
\nonumber\\
&=& 
\frac{\alpha_s^2}{4 \varepsilon}\,
\left [\, 1-\ \frac{1}{2h}\, \ln \left (\frac{1+h}{1-h} \right ) \right]\, ,
\label{pdp}
\eea
with  the
function  $h$ defined in Eq.~(\ref{hh}).
This is the general expression for the pole part of $\I^{(8\to 1)}$.
A short calculation shows that the lowest-order expansion of Eq.\ (\ref{pdp})
reproduces the result of Ref.\ \cite{fact2}, Eq.\ (\ref{svq}) given above
for the infrared pole at order $v^2$ (Electric dipole).
As anticipated above, our result is 
independent of the direction of the
light-like vector $l^\mu$ that defines the integrals.

To derive an equivalent form  in terms of the relative 
velocity, we use $P=P_1+P_2$ and $2q = P_1 -P_2$ and recall 
that we identify the relative velocity with
the heavy quark velocity in the pair center of mass: $\vec{v} = \vec{q}/E^*$, 
where $2E^*$ is the total energy of the heavy 
quark pair in this frame. In
these terms, we may replace the quantity $h$ of Eq.\ (\ref{pdp})
with an explicit function of velocity,
\bea
\I^{(8\to 1)}~=~ \frac{~\alpha_s^2}{4 \varepsilon} \; 
\left [1-\frac{1}{ 2f(|\vec{v}|)}\; 
\ln \left[\frac{1+ f(|\vec{v}|)}{1- f(|\vec{v}|)}\right]\, \right ]\, ,
\label{sl13}
\eea
where $v = |\vec {v} |$, and the function $f(v)$  is given by 
\bea
f(v) = \frac{2v}{1+v^2}\, . 
\label{sl14}
\eea
Equation (\ref{sl13}) is the general result at NNLO for the 
infrared pole.
As Eqs.\ (\ref{sl13}) and
(\ref{sl14}) show, the infrared 
term 
is independent of the eikonal momentum $l$
for finite $v$. 
Hence
the infrared pole structure at NNLO in $\alpha_s$ 
is consistent with factorization, which at this order 
in $\alpha_s$ is valid to all
orders in the relative velocity $v$ in the heavy quarkonium system. 

\section{Conclusions }

Equations (\ref{sl13}) and (\ref{sl14})  provides a remarkably compact expression for the 
single pole in the infrared factor ${\cal I}^{(8\to 1)}$ of Eq.\ (\ref{Edef}) at NNLO 
for finite relative velocity $v$, or equivalently, expanded to all orders in 
$v$. This result
is independent of the direction $l^\mu$ of the octet Wilson line, and hence is
consistent with NRQCD factorization, as discussed in connection with Eq.\ (\ref{schemefact}).
As shown in Ref.\  \cite{fact2}, the absence of $l$-dependence in
gauge-completed  matrix elements enables them to 
match all NNLO infrared divergent corrections to 
multijet cross sections.  All the arguments in that reference  for 
matching at order $v^2$ apply here
to all orders in $v$.

Although limited to NNLO, our result suggests that the decoupling of
light parton dynamics from heavy quark pair production is robust in
perturbation theory at the level of infrared divergences.  
Evidently, at NNLO, while a light-like, energetic parton
can resolve the color structure of a heavy quark pair, it does
so in a way that is independent of the 
direction of the
relative motion of the pair.
This suggests to us that the results derived here may generalize
to higher orders in soft gluon exchange.

In closing we note that there are, of course, many terms in a general NRQCD velocity expansion that
are not given by the eikonal approximation, in particular those that
deal with spin.  The eikonal approximation, however, does reproduce
infrared divergences as they appear in perturbative calculations of
heavy quark production.   Clearly, an extension of this analysis to
higher orders in 
the coupling will be of interest.

\subsection*{Acknowledgments}

This work
was supported in part by the National Science Foundation, grants
PHY-0354776 and PHY-0354822, and in part by the US
Department of Energy under Grant No.~DE-FG02-87ER-40371.

\end{document}